\documentclass[12pt]{article}
\usepackage{epsfig,amsfonts,amssymb}
\usepackage{hyperref}
\pdfoutput=1

\usepackage{cite}
\usepackage{cite}

\usepackage{comment}

\def\ZZZ{{\hbox{ Z\kern-1.6mm Z}}}
\def\RRR{{\hbox{ R\kern-2.4mm R}}}
\def\CCC{{\hbox{ C\kern-2.0mm C}}}
\def\zzz{{\hbox{z\kern-1mm z}}}

\newcommand{\qeq}{{\hbox{=\kern-2.3mm ? \kern.5mm }}}
\renewcommand{\qeq}{=}

\newcommand{\be}{\begin{equation}}
\newcommand{\ee}{\end{equation}}
\newcommand{\ben}{\begin{eqnarray}\displaystyle}
\newcommand{\een}{\end{eqnarray}}

\newcommand{\refb}[1]{(\ref{#1})}
\newcommand{\p}{\partial}
\newcommand{\sectiono}[1]{\section{#1}\setcounter{equation}{0}}

\def\one{{\hbox{ 1\kern-.8mm l}}}
\def\zero{{\hbox{ 0\kern-1.5mm 0}}}

\newcommand{\bea}[1]{\begin{eqnarray}\label{#1} }
\newcommand{\eea}{\end{eqnarray}}

\newcommand{\eqref}{\refb}

\usepackage{bm}
\usepackage[table]{xcolor}

\begin{document}

\baselineskip 24pt

\begin{center}

{\Large \bf String Theory in Rolling Tachyon Vacuum}

\end{center}

\vskip .6cm
\medskip

\vspace*{4.0ex}

\baselineskip=18pt

\centerline{\large \rm Ashoke Sen}

\vspace*{4.0ex}

\centerline{\large \it International Centre for Theoretical Sciences - TIFR 
}
\centerline{\large \it  Bengaluru - 560089, India}

%\centerline{\large \it ~$^c$Homi Bhabha National Institute}
%\centerline{\large \it Training School Complex, Anushakti Nagar,
%    Mumbai 400085, India}

\vspace*{1.0ex}
\centerline{\small E-mail:  ashoke.sen@icts.res.in}

\vspace*{5.0ex}

\centerline{\bf Abstract} \bigskip

We suggest that the universe filled with unstable D-branes in their rolling tachyon
vacuum state, described by periodic arrays of D-instantons along the imaginary time
direction, may be a natural background for formulating string theory. While the presence
of these D-instanton arrays do not affect the usual perturbative closed string amplitudes,
the open string degrees of freedom on the instanton may be used to create  the regular
D-branes via a series of marginal deformations, thereby describing D-branes as regular
classical solutions in the theory. 
Furthermore, we argue that a combination of the open string degrees of freedom in the
rolling tachyon vacuum is set equal to time by the equations of motion, and hence this 
combination could provide an intrinsic definition of time in the
theory. We illustrate these observations using the example of two dimensional string theory.

\vfill \eject

\tableofcontents

\sectiono{Introduction}

Rolling tachyon configurations describe time dependent solutions in string theory 
describing the classical decay of unstable D-brane 
systems\cite{0203211,0203265}.
They are characterized
by their energy (density) which can take arbitrary values between the energy (density) of
the original D-brane and zero.
The zero energy configurations can be called the rolling tachyon vacuum and will be the
subject of discussion in this paper.

The analytically continued
rolling tachyon solution to Euclidean time is described by an exactly marginal deformation 
of the boundary conformal field theory describing the original
D-brane.  As a simple consequence of this relation, one can
show that the rolling tachyon vacuum on a D0-brane 
in the bosonic string theory is described by a
periodic array
of D-instantons along the imaginary time axis. In the superstring or type 0 string theory the
analogous configuration is a periodic array of instantons and anti-instantons along the
imaginary time axis. In both cases the solution has zero energy. We can
place arbitrary number of such configurations in the system, still maintaining
vanishing energy. 
Furthermore, one can argue that perturbative closed string amplitudes
are not affected by the presence of these D-instanton arrays along the imaginary
time direction.  One could also consider generalization of this configuration where we
replace the periodic array of D-instantons by a periodic array of Euclidean D$(p-1)$-branes
describing rolling tachyon vacuum on an unstable D$p$-brane system, carrying vanishing
energy density.

In this paper we suggest that vacuum with such D-instanton and / or 
Euclidean D$(p-1)$-brane arrays may
be a more natural background for formulating string theories.
There are several motivations for this suggestion that we shall list below.

\begin{enumerate}
\item 
D-branes are natural objects in string theory, but they do not arise as regular
classical solutions in closed string theory since the former have tension of order $1/g_s$
while the latter have  tension proportional
to $1/g_s^2$, $g_s$ being the coupling constant of the closed string theory.
On the other hand, if we start with a vacuum containing a dense set of D-instanton arrays, 
then we can 
reach  classical configuration of D-branes
and anti-D-branes of different types, 
including time dependent rolling tachyon configurations with arbitrary
energy, by a series of marginal deformations. Since marginal deformations generate a
family of classical solutions, this shows that once we include the open string degrees of
freedom on the D-instanton arrays in the formulation of the theory, 
D-branes can be regarded as regular classical 
solutions in the theory.

In the past various approaches have been suggested for getting different D-branes as
classical solutions to the open string field theory on a specific D-brane
(see {\it e.g.} \cite{1406.3021,1909.11675} and references therein). 
These constructions
use the full power
of string field theory developed {\it e.g.} in \cite{0511286,1009.6185}. 
In contrast, the effect of marginal 
deformations can be studied using conformal perturbation theory and
hence may provide a more tractable description of D-branes as classical solutions.
\item D-instantons are known to give non-perturbative contribution to string amplitudes.
The construction of D-branes from periodic arrays of D-instantons described above does not
include isolated D-instantons of the kind that contribute to string amplitudes.
We show that in the presence of rolling tachyon vacua, there is a natural origin of the
D-instanton contribution to a string amplitude, -- instead of arising from a new saddle
point, it simply arises from a different choice
of time integration contour that threads one of the instantons that is already
present in the background.

\item  Another motivation for this proposal is to get a better understanding of the
tachyon effective field theory on unstable D$p$-branes. It is a theory of a scalar field $T$ 
with action
\be\label{etach1}
- \int d^{p+1} x \, V(T) \sqrt{1+\eta^{\mu\nu} \p_\mu T \p_\nu T}\, ,
\ee
where the tachyon potential $V(T)$ has a maximum at $T=0$ and approaches zero
at its minimum. This action was
originally proposed as a generalization of the Dirac-Born-Infeld
action for including the tachyon\cite{0003122,0003221}.
It was later found that if the minimum of $V(T)$ is taken at $T=\infty$, then
the late time dynamics of the
tachyon field in this effective field theory describes correctly the physics of rolling
tachyons\cite{0204143,0205098,0205085}.
In particular at late time the system reduces to pressureless matter, in agreement with
the result of boundary state analysis of the rolling tachyon solution in conformal field theory.
This theory is of interest since the equations of
motion of  the tachyon effective field theory sets the field $T$ to be equal to
time $t$ at large $t$, and hence the field $T$ may serve as an intrinsic definition
of time in the theory\cite{0209122}.
However this effective field theory was not derived from first principles. One hopes that
by studying the dynamics of open strings in the rolling tachyon vacuum,
we may be able to derive the tachyon effective field theory from first principles.

While in this paper we shall not be able to give a full derivation of the action 
\refb{etach1}, we argue  that there is a change of variable that maps spatially
homogeneous open string field configurations on the Euclidean D$(p-1)$-brane
arrays to a `time of flight' variable $\tau$ that is set equal to $t$ by the equations of
motion. Therefore it is natural to identify this to the field $T$ for large $T$ 
which is also set equal to $t$ by the equations of motion.  It
will be interesting to extend this to the full space-time dependent field configurations.

\end{enumerate}

The rest of the paper is organized as follows. In section \ref{sboson} we discuss
the basic idea in the context of rolling tachyon
vacuum in the bosonic string theory. In section \ref{sii} we generalize this to superstring
theories. In section \ref{s2d} we show how many of these ideas are realized in the context of
two dimensional bosonic and type 0B string theories for which there are dual free fermion
descriptions.  We also discuss the emergence of time from the rolling tachyon field.
We end in section \ref{send} with a discussion on possible future applications
of this approach.

\sectiono{Bosonic string theory} \label{sboson}

We shall begin our analysis with a D0-brane of bosonic string theory but 
the analysis below can be 
straightforwardly extended to the case of D$p$-branes by replacing the term
energy by energy density. The D0-brane has a tachyonic mode of
mass$^2=-1$ in the $\alpha'=1$ unit. The rolling tachyon configuration corresponds
to a time dependent background $T=\lambda\cosh(x^0)$ for real
parameter $\lambda$\cite{0203211,0203265}.
This describes an exactly marginal deformation of the original
D0-brane, as can be seen by noting that in the Euclidean time $x=ix^0$ this deformation
corresponds to the world-sheet perturbation 
$\lambda\cos(X)$. One can compute
the energy associated with this configuration by examining the boundary state, with the
result\cite{0203211,0203265,2310.17895}\footnote{It has been argued in 
\cite{0303139} that the rolling tachyon
solutions may eventually decay to closed string states and  that the total decay rate diverges
when we sum over all final state closed strings. A similar divergence exists in two dimensional
string theory, but  it was argued in \cite{0305159} that this phenomenon represents the 
ability to represent
D0-branes as coherent state of closed strings. It is not clear if a similar interpretation exists
in the critical string theory. Some discussion of this can be found in \cite{0410103}.}
\be
E=m_{D0}\cos^2(\pi\lambda)\, .
\ee 
For $\lambda=0$ this coincides with the D0-brane
mass $m_{D0}$ while for $\lambda=1/2$ this vanishes, corresponding to the vacuum of
the rolling tachyon configuration. For $0\le\lambda\le 1/2$ we have $m_{D0}\ge E\ge 0$,
describing a continuous family of 
solutions of different energies. Physically this
represents a motion of the tachyon where the tachyon comes from $\infty$ with energy
$E\le m_{D0}$, hits the potential barrier at $x^0=0$ and then returns back to infinity. One
can add a second parameter to this continuous family by shifting the time coordinate $x^0$
by a constant that changes the time at which the tachyon hits the potential 
barrier\cite{2310.17895}.

\subsection{Rolling tachyon vacuum} 

Our interest will be in the  vacuum solution $\lambda=1/2$ where $E$ vanishes. 
For Euclidean target space, the
conformal field theory at  $\lambda=1/2$ corresponds to a periodic array
of D-instantons placed at\cite{9902105,9402113,9404008,9811237}
\be 
x = (2n+1) \pi, \qquad n\in \ZZZ\, .
\ee
In the Lorentzian picture this will correspond to placing an array of D-instantons along
the imaginary time axis at
\be\label{e2.3}
x^0= (2n+1) i\pi, \qquad n\in \ZZZ\, .
\ee
We shall use the short form `D-instanton array' to describe such configuration of
D-instantons. Some remarks about this solution are in order.
\begin{enumerate}
\item
Even though the instantons are placed along the imaginary time axis, the configuration
gives real solution since the instantons are placed at complex conjugate pairs
of points.
\item The solution has zero energy but infinite Euclidean
action. Indeed the action associated with this
solution is independent of $\lambda$ and coincides with that of a D0-brane, since the
two solutions are related by a marginal deformation. For this reason this solution should be
thought of as a zero energy 
state of the theory instead of a finite action 
saddle point of the Euclidean path integral. 
\item The use of imaginary time is just a tool for
representing the solution. As an analogy, we can consider a classical field 
configuration,  expressed as
\be\label{e2.4}
\sum_{n\in\mathbb{Z}} F(x^0 - (2n+1)i\pi, \vec x)\, ,
\ee
for some function $F$ satisfying $\overline{F(x)}=F(\bar x)$. In that case \refb{e2.4} is real
for real $x^0$ and we do not need to move to the complex $x^0$ plane for physical
interpretation of the solution. 

\end{enumerate}

\subsection{Marginal deformations away from the vacuum}

Now we can address the question of whether starting from this solution we can get
rolling tachyon solutions of arbitrary energy. This requires reversing the marginal
deformation that produced this solution. The easiest way to describe this is
to regard the Euclidean time coordinate as a compact direction of radius 1 and make a T-duality
transformation. This maps the D-instanton to an euclidean
D0-brane spread along the dual Euclidean
time circle labelled by $y$.
We can now get a family of
rolling tachyon solutions by switching on deformations proportional
to $\cos(Y)$.
For appropriate
value of the deformation parameter the T-dual picture will produce a D-instanton which,
in the original description, corresponds to the original D0-brane with energy $m_{D0}$.
For intermediate values of the deformation parameter, we shall get rolling tachyon solutions
with energy $<m_{D0}$.

In the original description before T-duality transformation, 
deforming by $\cos(Y)=(e^{iY}+e^{-iY})/2$ amounts to
deforming by the vertex operator of the
winding mode of the string connecting the D-instanton to its image under
$2\pi$ translations and $-2\pi$ translations. In the non-compact theory with periodic array
of instantons, this will correspond to turning on background open strings stretched from
the $n$-th instanton to the $(n+1)$-th instanton for all $n$ simultaneously, as well as the 
open strings stretched from
the $(n+1)$-th instanton to the $n$-th instanton for all $n$ simultaneously. 

There are other marginal deformations associated with moving the D-instantons, either along
the spatial directions or along the imaginary time direction. All such configurations will have
zero energy density as long as no D-instanton lies on the real time axis. Note
however that once we destroy the periodicity along the imaginary time direction, we can no
longer switch on the winding mode deformations discussed earlier. The interpretation of 
this phenomenon 
in the context of two dimensional string theory will be discussed in section \ref{spap}. 
It is however not clear from this interpretation whether such deformations generate 
valid classical solutions in the Lorentzian signature. To avoid
dealing with such deformations, we shall regard the instantons as the $\lambda\to 1/2$ limit
of the solutions for generic $\lambda$ and only consider those marginal deformations
that respect the periodicity along the Euclidean time direction.

\subsection{Perturbative closed string amplitudes} \label{spert}

We shall now examine the fate of the usual perturbative closed string amplitudes in this
new vacuum.  In the absence of background D-branes the amplitude is expressed as a sum
over the contributions from Riemann surfaces without boundaries. 
In the presence of background D-branes, we need to include additional
Riemann surfaces with boundaries where we put the boundary condition relevant to the
background D-brane. Each such boundary can be replaced by an external closed string
source given by the boundary state associated with the D-brane. We shall now
argue that the boundary state associated with the rolling tachyon vacuum vanishes and
hence the perturbative closed string amplitudes are not affected by the presence of the
D-instanton arrays. 

The simplest argument is that since the D-instantons are localized
along $x^0=(2n+1)\pi i$, the boundary state is also localized at these points and hence
vanishes on the real $x^0$ axis. 
A more detailed argument is as follows. Since the D-instanton boundary state is given by
a combination of oscillators acting on a specific momentum carrying ground state, it is
sufficient to show that the ground state vanishes. For this we shall consider the case of
general $\lambda$. In that case the ground state is given by $F(X^0)(0)|0\rangle$ 
where\cite{0203211},
\be
F(x^0) = \sum_{n\in \mathbb{Z}} (-1)^n \sin^n(\pi\lambda) e^{-n x^0} \, .
\ee
This can be evaluated by dividing the sum into $n\ge 0$ and $n\le -1$ sets. They are
convergent respectively in the $x^0 > 0$ and $x^0<0$ domains. After carrying out these
sums and then continuing the results in the whole complex plane we get
\be\label{efx0exp}
F(x^0)={1\over 1 + e^{-x^0} \sin(\pi\lambda)} - {e^{x^0}\over 1 +e^{x^0} \sin (\pi\lambda) } \, .
\ee
For $\lambda=1/2$, this vanishes, confirming the vanishing
of the boundary state. 
This shows that the perturbative closed
string amplitudes in the new vacuum remain the same as in the usual vacuum, being given
by sum over contributions from Riemann surfaces without boundaries.

It may seem surprising that the boundary state describing the rolling tachyon vacuum 
vanishes even though the Euclidean
D-instantons can give non-vanishing contribution to the string
amplitudes and hence must have non-vanishing boundary state. The difference can be traced to
the difference in the choice of the time integration contour in the two
computations, {\it e.g.} while computing the Fourier transform
of the boundary state. For the rolling tachyon solution we take the time integration contour to
run along the real axis, while for computing D-instanton contribution to the amplitudes,
we take the time integral to run along the imaginary axis since we work in the Euclidean
space.
The difference in the result of integration over these two contours can be traced to
the fact that
\refb{efx0exp} has poles at $x^0=\pm\ln\sin(\pi\lambda) + (2n+1)i\pi$ and the integration
contour crosses these poles when we Wick rotate the real $x^0$-axis to the imaginary 
$x^0$ axis.\footnote{Such dependence of the momentum space boundary state on the
choice of $x^0$ integration contour has also been studied in \cite{0303139}.}
The residues at these poles precisely account for the difference between the 
integrals of \refb{efx0exp} along the real and imaginary time axes.
In particular, in the $\lambda\to 1/2$ limit,  these poles approach the
points $(2n+1)i\pi$ from opposite sides of the imaginary axis with opposite residues,
producing the $\delta$-functions at $(2n+1)i\pi$
where the D-instantons are  located.

Since the rolling tachyon vacuum does not affect perturbative closed string amplitudes, we
expect that this should be equivalent to the ordinary closed string vacuum. This will be
illustrated in section \ref{s2d} in the context of two dimensional string theory.

\subsection{Rolling tachyon vacuum to general D-branes} \label{sdbranes}

We can construct more general D-brane configurations by starting with multiple
D-instanton arrays. In particular, if we start with a dense\footnote{Having a density of order
unity in string units is sufficient to create higher dimensional branes via marginal
deformation.} set of such arrays placed at
different points in space, then by switching on independent winding deformation
along the imaginary time direction on each of D-instanton 
arrays we can produce configuration
of tachyon matter with different energy densities
in different spatial regions. 
We could also move these configurations relative to each other in real time directions
to construct superposition of time shifted rolling tachyon configurations.

We can also construct higher dimensional D-branes from such configurations. 
Let us choose a $p$-dimensional `square' lattice of
lattice spacing $2\pi$ along the spatial directions and 
place a D-instanton array at each of the lattice points. 
This will still be a zero energy configuration
since each array has zero energy. We can now switch on winding deformations of the
kind described earlier, but along the $p$ spatial directions. For appropriate value
of the deformation parameter it will convert the Dirichlet boundary condition along those
$p$ directions into Neumann boundary condition, producing an imaginary time 
periodic array
of euclidean D-$(p-1)$ branes.
This can be
interpreted as the rolling tachyon vacuum on a  D$p$-brane. We can now switch on
winding deformations along the Euclidean time direction to produce rolling tachyon solution
on the D$p$-brane with finite energy density. In appropriate limit of this deformation 
parameter we shall recover the static D$p$-brane configuration without any rolling tachyon
deformation.

Note that
we could also start with any number of arrays of space filling  Euclidean D-branes located
at $x^0=(2n+1)i\pi$,
produce D-instanton arrays at each point on the spatial
lattice by switching on deformations
proportional to $\cos(X^i)$ 
along the spatial directions and then redistribute these arrays in any way we like
by moving them along the spatial directions. 
All of these represent marginal deformations within zero energy density
subspace. More generally the vacuum could
contain different types of euclidean D-brane arrays extending along different spatial directions
but lying at the points \refb{e2.3} along the Euclidean time direction.

\subsection{D-instantons} \label{sdinst}

\def\figtwo{

\def\JPicScale{0.8}
\ifx\JPicScale\undefined\def\JPicScale{1}\fi
\unitlength \JPicScale mm
\begin{picture}(70,70)(0,0)
\linethickness{0.3mm}
\put(40,30){\line(0,1){20}}
\linethickness{0.3mm}
\put(10,50){\line(1,0){30}}
\linethickness{0.3mm}
\put(40,30){\line(1,0){30}}

\put(39,40){\makebox(0,0)[cc]{$\bullet$}}
\put(41,40){\makebox(0,0)[cc]{$\bullet$}}

\put(40,45){\makebox(0,0)[cc]{$\downarrow$}}
\put(25,49.6){\makebox(0,0)[cc]{$\rightarrow$}}
\put(55,29.6){\makebox(0,0)[cc]{$\rightarrow$}}

\end{picture}

}

\begin{figure}
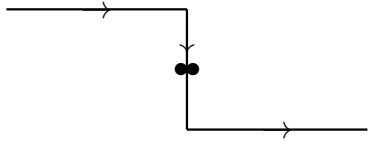


\begin{center}

~

\vskip -.7in

\figtwo

\vskip -.7 in

\caption{This figure shows the time integration contour associated with D-instanton
amplitudes. The D-instanton is shown by a pair of dots on two sides of the contour
to indicate that it should be regarded as a pair of poles with opposite
residues as described in section \ref{spert}.} \label{figtwo}

\end{center}

\end{figure}

One feature of string theory that does not seem to be captured by the analysis in
section \ref{sdbranes} is the usual D-instanton contribution to string amplitudes. 
Conventionally, D-instanton contribution to a string amplitude is described as a
contribution from a new saddle point of the Euclidean path integral. However, the
procedure described in section \ref{sdbranes} for constructing D-brane solutions via
marginal deformation does not apply to isolated
D-instantons themselves. Instead of trying
to look for new saddle points, 
we tentatively identify the D-instanton contribution as  a choice of
time integration contour that runs along the real axis from $-\infty$ to 0, travels a
distance $2\pi$ along the
imaginary axis and then continues along the real axis towards $\infty$. The
boundary state associated with this time integration contour will get contribution from a
single D-instanton that the contour crosses. 
Again, it is best to describe the
instanton as the $\lambda\to 0$ limit of the rolling tachyon solution, so that the 
instanton can
be thought of as a pair of poles on two sides of the imaginary axis with opposite residues.
This has been shown in Fig.~\ref{figtwo} where
the $\bullet\bullet$ denotes the instanton that the contour threads. 
This gives the instanton contribution at a fixed space-time 
coordinate, but the divergence
in the annulus partition function will force us to integrate over the D-instanton 
position\cite{2101.08566,2104.11109}.
We can carry out this integration along the real time (and space) directions, producing the
usual energy-momentum conserving delta functions in the amplitude.

This will
be discussed further in section \ref{shole} in the context of two dimensional string theory.

\sectiono{Superstring theory} \label{sii}

The analysis for superstring theory is similar. The unstable D-brane system under 
consideration is either a non-BPS D-brane\cite{9806155,9808141,9809111,9904207}
or a coincident D-brane - anti-D-brane system.
Let us again focus on D0-branes for definiteness but the generalization to D$p$-branes
is straightforward. The open string spectrum
has a tachyon of mass$^2=-1/2$ in
the $\alpha'=1$ unit. The rolling tachyon solution is obtained by switching on a tachyon
background of the form $\lambda\cosh(x^0/\sqrt 2)$, and has energy $M\cos^2(\lambda\pi)$
where $M$ is the mass of the non-BPS D0-brane or the total mass of D0 and $\bar{\rm D}$0
branes, as the case may be.  The rolling tachyon vacuum corresponds to the $\lambda=1/2$
configuration where the energy vanishes. By going to the Euclidean theory one can show that this 
has a description as a system of (anti-)D-instantons along the imaginary time axis,
placed at\cite{9808141,9904207,9903123},
\be\label{e3.1}
x^0 = {1\over \sqrt 2} \pi i \left(2n+1\right)\, , \qquad n\in \ZZZ\, .
\ee
If the starting configuration is a  non-BPS D0-brane then we place
a BPS D-instanton for odd $n$ and BPS anti-D-instanton for even $n$ (or vice versa if we
consider the $\lambda=-1/2$ configuration), 
while if the starting configuration is a 
D0-$\bar{\rm D}$0 pair then we place a non-BPS D-instanton
at every point corresponding to integer $n$.
We call this the D-instanton array in superstring theory.  Note that
this is a real solution since under complex conjugation and exchange of D-instanton and
anti-D-instanton, the solution remains invariant\cite{2012.00041}. The rest of the
discussion is identical to that in the case of bosonic string theory. In particular we can 
construct rolling tachyon configurations of finite energy from this rolling tachyon
vacuum solution by switching on winding deformations
along the imaginary time direction. Also by taking a dense set of such arrays (of both
kinds, one where the instantons are placed at odd $n$ and the other where the
anti-instantons are placed at odd $n$) we can produce other D-branes as well as rolling
tachyon solutions on these branes via marginal deformation.

Note that using marginal deformations of the kind described above, we can only reach
charge neutral configurations,
 -- these could be either non-BPS D-branes or coincident
D-brane $\bar{\rm D}$-brane
pairs.  However once we have created a brane - anti-brane pair, we can
use marginal deformations to separate them along the transverse directions, possibly
moving the unwanted branes to infinity.
This way we can construct arbitrary configuration of D-branes with compensating charges
at infinity.

\sectiono{Insights from two dimensional string theory} \label{s2d}

Two dimensional string theory can be regarded as a critical string theory whose
world-sheet matter theory contains a free scalar field associated with the time direction and
a Liouville field with central charge 25, labelling the space 
direction\cite{dj,sw,gk,kr,0304224,0305159,1705.07151,1907.07688}.
A different version of this theory is a type 0B theory whose world-sheet matter theory has
a free scalar labelling the time direction and its supersymmetric partner fermion and
a super-Liouville theory of $\hat c=9$ labelling the space 
direction\cite{0307083,0307195,2201.05621,2204.01747}. 
Both theories have
a dual description 
as a theory of free fermions in an inverted harmonic oscillator potential with
Hamiltonian 
\be\label{eham}
H= {1\over 2} \, p^2  -{1\over 2} \, q^2 +\mu\, ,
\ee 
where $\mu=g_s^{-1}$ is the inverse string coupling.
For the dual of the bosonic string theory the
vacuum corresponds to filled fermi sea up to zero energy 
on the $q>0$ side leaving the $q<0$ side
empty, -- a precise description of this can be found in \cite{1907.07688}.
For the dual of the type 0B string theory the
vacuum corresponds to filled fermi sea up to zero energy on both sides.
In both theories the closed strings are represented as fermion hole pairs,
while the rolling tachyon solution corresponds to a single fermion with energy above the
fermi sea.

In this section we shall examine how the properties of
classical solutions, describing the motion of a free
fermion in the inverted harmonic oscillator potential given in \refb{eham}, match with the
properties of rolling tachyon solutions in the two dimensional string theory.
The insights gained from this have
direct bearing on the rolling tachyon solutions in the critical string theory as well, since
the rolling tachyon solution in open bosonic 
string field theory belongs to the `semi-universal sector',
involving world-sheet states created from the SL(2,R) invariant vacua by the action of the
$X^0$ and ghost operators and the Virasoro generators of the remaining $c=25$ conformal
field theory but does not involve states built from non-trivial primaries of the $c=25$
CFT. Therefore the rolling tachyon solutions in the two dimensional bosonic string theory are
identical to those on any D$p$-brane 
in any compactification of critical bosonic string theory. Similarly,
rolling tachyon solutions in two dimensional type 0B string theory are identical to those in
any compactification of critical type 0A/0B, type I or type IIA/IIB string theory.

\subsection{Rolling tachyon vacuum}

In the free fermion description, a  D0-brane in the rolling tachyon vacuum
corresponds to the addition of a zero energy fermion to the fermi sea. Classically the
fermion follows the trajectory $q=\sqrt{2\mu} \cosh x^0$. The description of this in the
Euclidean time $x$ is $q=\sqrt{2\mu}\, \cos x$ which precisely corresponds to D-instanton
solution in this theory. In the bosonic string theory, a single D-instanton 
corresponds to one full oscillation in $x$ and hence the
solution $q=\sqrt{2\mu} \cos x$ indeed represents an array of D-instantons along the imaginary time direction placed
at interval $2\pi$. In the type 0B interpretation the D-instanton represents half an oscillation
from one turning point to another and the anti-instanton represents the other half of the
oscillation where the motion is in the opposite direction.
Therefore the solution represents
an array of
instantons and anti-instantons as in section \ref{sii}.
The apparent mismatch
between the period $2\pi$ in the matrix model and the period $2\pi\sqrt 2$ in \refb{e3.1} can be
traced to the fact that \refb{e3.1} is given in the $\alpha'=1$ unit, while
the free fermion Hamiltonian used here describes type 0B string theory
with $\alpha'=1/2$ and bosonic string theory with $\alpha'=1$.

Since the rolling tachyon vacuum carries zero energy, one expects that in some sense
this should be equivalent to the ordinary vacuum. This is also best illustrated in the 
fermionic description, since adding a
fermion at the fermi level
has no effect on the vacuum. However, having the extra  degree of freedom allows us to give a
simple description of the states where a fermion carries non-zero energy, by simply
deforming the solution with zero energy fermion to a solution with non-zero energy fermion.
This describes finite energy rolling tachyon solution on a D0-brane.
In contrast, if we only had access to closed string degrees of freedom, corresponding to
fermion hole pair excitations, then the description of single fermion excitations will be
complicated and involve infinite number of closed string states. Of course at the full
non-perturbative level we should not include both the closed and open string degrees of
freedom in the fundamental description of the theory, but at the perturbative level having
both degrees of freedom may be useful {\it e.g.} in studying scattering of closed strings and
D0-branes.

\subsection{Periodic vs aperiodic D-instanton configurations} \label{spap}

Note that 
the free fermion trajectory described by the solution $q=\sqrt{2(\mu-E)}\cosh x^0$, when
analytically continued to Euclidean time,
is always periodic with period $2\pi$ for $E>0$. This fits in well with the results
in the dual string theory that the finite energy configurations, corresponding to $\lambda<1/2$,
are always periodic along the Euclidean time direction with $\lambda$-independent 
period. However, exactly
at $\lambda=1/2$, when the energy of the D0-brane vanishes, the solution in string theory
admits marginal deformations that take us away from periodic configuration. This is related to
the existence of single D-instanton solutions without having to have periodic array. This suggests
that in the dual free fermion description we must also have such solutions. 
We shall now give a qualitative description of how this might arise.

As already described above, in the free fermion description we have filled fermi sea 
on one or both sides of the potential for $E<0$. This means in the classical phase space,
the region outside the hyperbola $p^2-q^2+2\mu\le 0$, $q\ge \sqrt{2\mu}$  
is excluded in the bosonic
string theory while the region outside the hyperbola $p^2-q^2+2\mu\le 0$ for either sign of $q$
is excluded in the type 0B string theory. We could regard this as an infinite repulsive potential
that arises when a phase space trajectory lies inside the forbidden region.

Now, the trajectory describing the instanton solution is in the Euclidean space where
$p=ip_E$ is purely imaginary and the trajectories take the form of circles:
$2E=2\mu -p_E^2 -q^2$.
However,
the turning points where $p_E$ vanishes are real points in the phase space. 
It is easy to see
that for the $E=0$ trajectory, the turning point at $p_E=0$, $q=\sqrt{2\mu}$ 
lies at the boundary of the excluded region in the bosonic string theory. 
Therefore the fermion
should feel infinite potential as it approaches this point and the trajectory ends
there without continuing further.\footnote{The infinite repulsive potential in the Lorentzian theory
corresponds to an infinite attractive potential in the Euclidean theory. Therefore in the Euclidean
description the zero energy particle travels along the trajectory of a usual harmonic oscillator
with potential $q^2/2-\mu$
and gets stuck at the wall at $q=\sqrt{2\mu}$ instead of getting reflected.}
As a result, in the phase space, instead of covering the circle
$p_E^2+q^2=2\mu$ infinite
number of times, as will be the case for a free fermion, the trajectory will cover the
circle only
once, approaching the point $p=0$, $q=\sqrt{2\mu}$ asymptotically at both ends. This will
describe a single D-instanton instead of a periodic array of D-instantons.

In type 0B theory, both the turning points $p=0$, $q=\pm\sqrt{2\mu}$ lie at the boundary
of the forbidden region and hence the fermion will feel infinite potential 
as it approaches these points. Therefore now the circular trajectory $p_E^2+q^2=2\mu$
is cut at two places and one has two independent phase space trajectories, one that
begins at $\sqrt{2\mu}$ and ends at $-\sqrt{2\mu}$ and the other one begins at
$-\sqrt{2\mu}$ and ends at $\sqrt{2\mu}$. These can be regarded as single D-instantons
and anti-D-instantons respectively.

\subsection{Time from rolling tachyon}  \label{stime}

A convenient set of phase space coordinates for describing the dynamics of a particle with
Hamiltonian \refb{eham} are obtained by
making a canonical transformation from $q,p$ to the time of flight and
energy variables:
\be \label{etimeofflight}
h = {1\over 2} (p^2-q^2+2\mu), \qquad \tau = \sinh^{-1} {p \over \sqrt{q^2 - p^2}}\, ,
\ee
so that we have $H=h$ and the equations of motion set $h$ to a constant and $\tau$ to $t$.
Note that this is a non-singular change of variables even as we approach the Fermi
surface $h=0$.
Since the same dynamics is supposed to be captured by open string field theory on an
unstable D0-brane, this implies that there must be a change of variable that maps some
combination of open string fields to the variable $\tau$ that is set equal to time by the
equations of motion. In fact, since the 
tachyon field of string field theory shows wild oscillation in time\cite{0207107},
it must be some combination of the tachyon and higher level fields that map to the
time of flight variable.

This has consequences beyond two dimensional string theory. As explained in the
introduction to this section, the rolling tachyon solution enjoys a semi-universality property, namely
that the components of the string field that are involved in the solution are obtained by
acting on the SL(2,R) invariant vacuum by operators constructed from the
time coordinate $X^0$, the ghost fields and the stress tensor of the $c=25$ Liouville theory,
but does not involve the non-trivial primaries of the Liouville theory. This means that
exactly the same solution exists in any (compactification of the) 
bosonic string theory whose time coordinate 
describes a $c=1$ CFT that is decoupled from the rest of the $c=25$ matter theory,
and hence the same change of variable will map the open string fields on the D0-brane
(or spatially homogeneous open string fields on a D$p$-brane)
to the variable $\tau$ that is set equal to time by its equation of motion. A similar argument
holds for the rolling tachyon solution on the D0-brane of type 0B string theory 
but the solution also involves the superconformal ghosts,
the superpartners of $X^0$ and the supercurrents of the $\hat c=9$ theory. This
solution  now describes the rolling tachyon solution on any unstable D-brane 
in any compactification
of type 0A/0B string theory, IIA/IIB string theory or type I string theory,
since in the subsector in which the rolling tachyon solution lies, the open
string field theories in these different string theories are identical. 
Therefore we can find a universal change of variable involving
the open string fields that maps an appropriate combination of the string
fields to the variable $\tau$.\footnote{Here universality means that in all compactifications of bosonic string
theory there is a universal combination of 
open string fields that is set equal to time by the equations of motion, 
and in all compactifications of type IIA/IIB, type 0A/0B and type I string theory, there is another
universal combination of 
open string fields that is set equal to the time by the equations of motion. 
} 
In particular, {\em 
if we begin with a space-filling brane system then $\tau$ describes a scalar field on the
brane and provides an intrinsic definition of time in the theory}.\footnote{If instead 
we consider the vacuum containing a
dense set of D-instanton arrays, we could take local average of the time of flight variables over
string scale volume and identify that as a course grained scalar field that plays the role
of time.}

The idea of using
scalar fields as time of course is not new; but the novelty of the present case is that we
can use the variable $\tau$ as the time coordinate even for zero energy where
physically the background is
time independent. In the context of tachyon effective field theory, this was discussed in
\cite{0209122}, but the effective field theory was not derived from first principles.

\subsection{Time independent tachyon vacuum}

In the discussion in section \ref{spap}, 
since the Euclidean phase space trajectories stop at $p_E=0$, $q=\sqrt{2\mu}$,
this point will describe a time independent solution of the equations of motion once we
take into account the repulsive  force due to the filled fermi sea. 
Using $p=ip_E$ we see that this also exists as a real solution in the Lorentzian theory.
One way to see the existence
of such a solution is to use the canonically conjugate pair of 
variables $(h,\tau)$ introduced in \refb{etimeofflight}.
To represent the exclusion of the phase space region below the Fermi sea, we can modify the
Hamiltonian $H=h$ to
\be\label{emodham}
H = \cases{h \ \hbox{for} \ h>0\cr \infty \ \hbox{for} \ h\le 0}\, .
\ee
This has been shown in Fig.~\ref{figone}.
If we now regulate the Hamiltonian by rounding off the sharp edge at $h=0$, we shall
get a sharp minimum of $H$ at $h=0$, as shown by the point $P$ in Fig.~\ref{figone}. 
The equations of motion now admit a time
independent solution
$h=0$, $\tau=$constant. This describes an infinite family of time independent solutions
labelled by the value of $\tau$, but
we could lift the degeneracy by making $H$ depend on $\tau$ near $h=0$, {\it e.g.} by taking
$H = h \{1-e^{-\alpha h (1+\tau^2)}\}$ so that in the $\alpha\to \infty$ limit we
get back \refb{emodham}. With this regulated Hamiltonian 
we get a time independent solution at $\tau=0$,
$h\simeq 0$ which translates to $(p,q)\simeq (0,\sqrt{2\mu})$.

\def\figone{

\def\JPicScale{0.8}
\ifx\JPicScale\undefined\def\JPicScale{1}\fi
\unitlength \JPicScale mm
\begin{picture}(70,70)(0,0)
\linethickness{0.7mm}
\put(30,30){\line(0,1){40}}
\linethickness{0.7mm}
\multiput(30,30)(0.12,0.12){333}{\line(1,0){0.12}}
\put(30,25){\makebox(0,0)[cc]{$P$}}

\put(20,50){\makebox(0,0)[cc]{$H\uparrow$}}

\put(50,25){\makebox(0,0)[cc]{$h\rightarrow$}}

\end{picture}

}

\begin{figure}
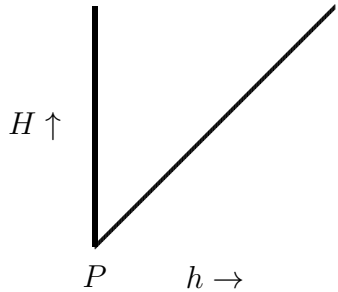


\begin{center}

\figone

\vskip -.7 in

\caption{Modification of the Hamiltonian $H$ as a function of $h$
due to the fermi pressure} \label{figone}

\end{center}

\end{figure}

It is natural to
conjecture that this describes the  time independent tachyon
vacuum solution studied in \cite{9911116,Kostelecky:1988ta,
9912249,0002211,0002237,0211012,0511286}. 
As a classical solution this looks very different from the
rolling tachyon vacuum, but in the quantum theory where the boundary between the
allowed and excluded regions becomes fuzzy, 
both should correspond to the  same 
zero energy quantum state.

\subsection{Holes and instantons} \label{shole}

The free fermion description of two dimensional string theory also has hole like excitations
that correspond to removing some fermions from below the Fermi level.
Some suggestions for what
the hole states could be in string theory have been made in \cite{0307195,0307221}.
This requires analytically continuing the
deformation parameter $\lambda$ to complex value and then switching the sign of the
boundary state. 
Open string spectrum in such a
system is difficult to study, and for this reason
there has not been
a detailed study of the underlying conformal field theory description of the hole states. 
However, if one is looking for the `hole vacuum' 
which corresponds to simply removing a zero energy fermion  
from the fermi sea,
one expects it to be given by a `subtraction' of the rolling tachyon vacuum. This 
should be described by
a boundary state that is negative of the boundary state of the D-instanton array
describing the rolling tachyon vacuum. If we call this an array of
`negative instantons' (distinct from anti-instantons for which only the RR sector
of the boundary state changes sign), then the negative instantons will have the
property that the open strings between two negative instantons will have spectrum
identical to that for open strings between two ordinary instantons whereas the
open strings between a positive instanton and a negative instanton will have their
spectrum of fermionic and bosonic states interchanged so that the annulus partition
function changes sign. It will be interesting to
explore whether this describes a sensible state in string theory. If it does, then by
the universality of the open string field theory, such configurations should also exist
in the critical string theories.

On the other hand, if
we begin with a vacuum containing one or more D-instanton arrays representing extra
fermions at zero energy, then the creation of a hole vacuum state 
can be described as removal of
one of these arrays, and no exotic D-branes will be necessary for describing this process.
For example, in two dimensional type 0B theory a zero energy fermion
on the right side of the potential corresponds to a D-instanton array with instantons at odd
site and anti-instantons on the even site and  a zero energy fermion
on the left side of the potential corresponds to a D-instanton array with instantons at even
site and anti-instantons on the odd site. Therefore if we want to describe a final state that
has an extra fermion on the left side and a hole on the right side compared to the initial state,
we need to shift
one of the arrays that had instantons at even site by half a period along the imaginary
axis so that the positions of instantons and anti-instantons get exchanged. 
This can be achieved by choosing  
the integration contour over time to run along the real axis from
$-\infty$ to some finite real time where a D-instanton array is located, 
then travel half a period along  the imaginary axis treading an instanton
and then continue parallel
to the real axis towards $\infty$. In this case on the incoming trajectory the solution will
represent a fermion on the right side of the potential and on the outgoing trajectory the
solution will represent a fermion on the left side of the potential.
This is precisely the D-instanton induced process as described in section \ref{sdinst}.

\sectiono{Discussion} \label{send}

In this paper we have suggested that in the formulation of string theory, besides the usual
closed string degrees of freedom, we could add the open string degrees of freedom living
on the rolling tachyon vacuum. We can then use marginal deformations in the open string
sector to describe any D-brane configuration with compensating charges at
infinity. These could either
be time independent configurations of regular D-branes or rolling tachyon configurations on
unstable D-branes. We also argued that there is a combination of the open string field
variables that is set equal to time by the equations of motion and hence could serve as the
intrinsic definition of time in the quantum theory. 

Drawing lessons from the two dimensional string theory, we can see that at a fundamental
level the open and closed string degrees of freedom should not be treated as independent
-- there the open string degrees of freedom are results of fermionization of closed string
degrees of freedom. Nevertheless, having both sets of degrees of freedom
is useful as an effective theory since this gives a simpler
description of finite energy D0-brane configurations, which otherwise need to be described
by a coherent state of closed strings. The same might be true in
critical string theory where the solution may be described as a coherent state of massive
closed string states describing the pressureless matter that one obtains from the rolling tachyon solution at late time\cite{0203211,0203265}.
Of course eventually the massive closed string states should decay into massless states but
that is a higher order effect in string coupling.
A more radical
point of view would be that we may be able to reformulate string theory purely in terms of the
open string degrees of freedom -- this will be the analog of fermionic formulation of the two
dimensional string theory where closed strings can be described as fermion hole pairs. However
to generalize this to the critical string theory we need a better understanding of the hole states
in string theory than what we have at present.

As already mentioned,
one of the motivations for this paper is  that this line of analysis may lead to
a better understanding of the tachyon effective field theory \refb{etach1}
used for describing
the rolling tachyon solution. 
So far we have had partial success in this front. A full success will establish that the
generic inhomogeneous
decay of an unstable D-brane system will  produce a system of non-rotating,
non-interacting dust at late time.
Using this result and 
earlier work of \cite{Kuchar:1990vy,9409001},
it was shown in \cite{0209122} that when
we couple the tachyon effective field theory to gravity, the Wheeler-deWitt equation
reduces to standard Schrodinger equation in quantum mechanics, with the tachyon field
itself serving as the definition of time. 
This could also gel with the recent emphasis on the role of
the observer in  quantizing gravity\cite{2206.10780,2308.03663}, with the 
tachyon dust providing a natural set of such observers in string theory. 

\bigskip

\noindent{\bf Acknowledgement:} 
This work was supported by the ICTS-Infosys Madhava 
Chair Professorship, the J. C. Bose fellowship of the Department of Science
and Technology, India
and the Department of Atomic Energy, Government of India, under project no. RTI4001.

\end{document}